# Electrical Detection of Spin Backflow from an Antiferromagnetic Insulator/Y$_3$Fe$_5$O$_{12}$ Interface

Weiwei Lin[1,*] and C. L. Chien[1,†]

*Department of Physics and Astronomy, Johns Hopkins University, Baltimore, Maryland 21218, USA*

**Abstract**

Spin Hall magnetoresistance (SMR) has been observed in Pt/NiO/Y$_3$Fe$_5$O$_{12}$ (YIG) heterostructures with characteristics very different from those in Pt/YIG. We show that the SMR in Pt/NiO/YIG strongly correlates with spin conductance, both sharing very strong temperature dependence due to antiferromagnetic magnons and spin fluctuation. This phenomenon indicates that spin current generated by spin Hall effect in the Pt transmits through the insulating NiO and is reflected from the NiO/YIG interface. Inverted SMR has been observed below a temperature which increases with the NiO thickness, suggesting spin-flip reflection from the antiferromagnetic NiO exchange coupled with the YIG.

[*]wlin@jhu.edu
[†]clchien@jhu.edu

Recent advents in spintronics have led to the exploitation of pure spin current, which efficiently transports spin angular momentum without accompanied by a charge current thus generating no Oersted field and less Joule heating [1-9]. Pure spin current phenomena, such as nonlocal spin injection [1,2], spin pumping [3,4], spin Hall effect (SHE) [5,6], inverse spin Hall effect (ISHE) [7,8], and spin Seebeck effect (SSE) [9], have been explored in heterostructures consisting of normal metals (NMs), ferromagnetic (FM) metals, ferromagnetic insulators (FMIs) [1-4,6,8], and very recently, also antiferromagnetic (AF) materials [10-20]. It has been recently observed using spin pumping and SSE that a thin antiferromagnetic insulator (AFI), such as NiO and CoO, when inserted between a NM layer and a ferrimagnetic insulator yttrium iron garnet (YIG = $Y_3Fe_5O_{12}$) as in NM/AFI/YIG, not only transmits but also enhances spin current by as much as one order of magnitude [12-15]. The spin current enhancement exhibits a maximum near the Néel temperature of the thin AF layer, highlighting the central role of spin fluctuations in the AF layer [15-18]. These attributes of AFs may facilitate new roles in pure spin current phenomena and devices, which thus far have largely excluded AF materials.

Thanks to the spin Hall magnetoresistance (SMR), first observed in NM/FMI structures such as Pt/YIG [21-27], the spin current reflection at the NM/FMI interface can be detected electrically. According to the SMR theory [23,24], spin current $\mathbf{J}_S^{SH}$ generated by the SHE in Pt is either reflected ($\mathbf{M} \parallel \boldsymbol{\sigma}$, where $\boldsymbol{\sigma}$ is spin current polarization) or absorbed ($\mathbf{M} \perp \boldsymbol{\sigma}$) at the Pt/YIG interface. Then, the reflected spin current $\mathbf{J}_S^r$ is converted to an additional charge current $\mathbf{J}_C^{ISH}$ due to the ISHE in Pt, where $\mathbf{J}_C^{ISH} \propto \mathbf{J}_S^r \times \boldsymbol{\sigma}$. Thus, it leads to a decrease of the measured resistance in Pt because the direction of $\mathbf{J}_C^{ISH}$ is parallel to that of the applied charge current $\mathbf{J}_C$ [23,24].

In this Letter, we show that SMR in NM/AFI/YIG heterostructures (NM = Pt or W, AFI = NiO or CoO) reveals spin current *reflection* from the AFI/FMI interface, as well as enhanced transmission through the AFI layer. Note that the SMR in NM/AFI/YIG quantifies *magnon* spin current reflection from an *AFI/FMI* interface, rather than spin current reflection from a NM/FMI interface as in the conventional SMR. Importantly, strong temperature dependence of the SMR in Pt/NiO/YIG reflects that of spin conductance, completely different from that in Pt/YIG. We have observed inverted SMR in the Pt/NiO/YIG at low temperatures, suggesting *spin-flip* reflection from the AF NiO exchange coupled with the YIG.

We used magnetron sputter to deposit thin films onto polished polycrystalline YIG substrates with 0.5 mm thickness via dc Ar sputtering for Pt and W, reactive Ar + $O_2$ sputtering for NiO and rf Ar sputtering for CoO at ambient temperature. X-ray diffraction shows all the layers are polycrystalline, as shown in Fig. 1(a). The films were patterned into 5 mm long Hall bar structures with 0.2 mm wide lines 1.5 mm apart by photo-lithography. As sketched in Fig. 1(b), the magnetoresistance (MR) of the wire was measured with current **I** in the long segment ($x$) and voltage measured at the two short segments. The measured resistance depends on the direction of the magnetization **M** of the underlying YIG as aligned by a magnetic field. In particular, with **M** in the film plane one measures longitudinal $R_\parallel$ (**M** along $x$ and $\parallel$ **I**) and transverse $R_T$ (**M** along y and $\perp$ **I**), and with **M** out of the film plane, perpendicular $R_\perp$ (**M** along z and $\perp$ **I**). The magnetic field **H** was applied in the $xy$, $yz$ and $zx$ planes with angles $\alpha$, $\beta$ and $\gamma$ relative to the $x$, $z$ and $x$ directions, respectively.

Figure 1(c) shows the MR of the Pt(3)/NiO(1)/YIG (the numbers in parentheses are thickness in nm) at temperature $T$ = 300 K with field along $x$ and $y$ axes for $R_\parallel$ and $R_T$, respectively and showing $R_\parallel > R_T$. In contrast, no MR is observed in Pt(3)/NiO(1)/SiO$_X$/Si, as shown in Fig. 1(c). This means that the MR measured in the Pt layer requires the presence of ferrimagnetic YIG. The angular scan in the $xy$ ($yz$) plane under 0.5 T field (enough to saturate the YIG magnetization) shows $\cos^2\alpha$ ($\cos^2\beta$) behavior in resistance, and the $\gamma$ scan in the $zx$ plane has no variation, as shown in Fig. 1(d). These results confirm $R_\perp \approx R_\parallel > R_T$ at $T$ = 300 K in the Pt/NiO/YIG, the same characteristics as the SMR in Pt/YIG [21-27]. SMR has the unique characteristics of $R_\perp \approx R_\parallel > R_T$, differing from all other known MR, such as the anisotropic magnetoresistance (AMR) in FM metal with $R_\parallel > R_T \approx R_\perp$. Note that the behavior of $R_\perp \approx R_\parallel$ is essential in establishing SMR and distinguishing from the AMR [25,26].

Bulk NiO has a Néel temperature $T_N$ of 535 K. However, $T_N$ of thin NiO films is much lower due to finite-size effects. For 1 nm NiO, $T_N$ is about 170 K [15] with no AF ordering in 1 nm NiO film at room temperature (RT). Spin-dependent scattering at the Pt/NiO interface can be excluded as the origin of the SMR observed in the Pt/NiO/YIG above the $T_N$ of the NiO layer. The SMR observed in the Pt/NiO/YIG above the $T_N$ of the NiO layer indicates that spin current generated by the SHE in the Pt transmits through the insulating NiO and is reflected (absorbed) at the NiO/YIG interface as **M** $\parallel$ **σ** (**M** $\perp$ **σ**), as sketched in Fig. 1(e). It means that the SMR of NM reveals the magnetization orientation of FMI even when separated by an insulating spacer because NiO transmits pure spin current.

Since the SMR is due to spin current transmission through NiO, it depends sensitively on the NiO thickness. Figure 2(a) shows the NiO thickness dependence of the SMR ratio $\Delta R/R_0 = (R_\parallel - R_T)/R_0$ in Pt(3)/NiO($t_{NiO}$)/YIG at RT, where $R_0$ is the zero field resistance. SMR is detectable only for $t_{NiO}$ less than about 3 nm, within which the $\Delta R/R_0$ value actually enhances with a peak at $t_{NiO}$ ~ 1 nm. The enhancement is about 2 in Pt(3)/NiO/YIG over that of Pt(3)/YIG, whereas an even larger enhancement of about 6 is observed in W(3)/CoO($t_{CoO}$)/YIG, with a maximum at $t_{CoO}$ ~ 1.4 nm. With increasing AFI layer thickness, the SMR in Pt/NiO/YIG and W/CoO/YIG eventually vanishes at $t_{NiO}$ > 3 nm and $t_{CoO}$ > 3.4 nm respectively. Note that the thickness dependence of the enhancement in SMR is very similar to the pure spin current enhancement in Pt/NiO/YIG recently observed by SSE in the same structure [15] due to the intimate relationship between spin current and SMR. One may notice that the AFI thickness for the SMR decay is smaller than that for the SSE measurement [15]. This is because in the SMR measurement, the spin current generated from the NM via the SHE transmits though the AFI layer, reflected from the AFI/FMI interface, transmits through the AFI layer again and is detected in the NM. The spin current makes a round trip passage through the AFI for the SMR measurement, while only a single way passage in the SSE measurement.

Temperature dependences of SMR in Pt/YIG and Pt/NiO/YIG are shown in Fig. 3(a). The $\Delta R/R_0$ in Pt/YIG shows weak $T$ dependence, whereas those of Pt/NiO/YIG show very strong $T$ dependences. The $\Delta R/R_0$ of Pt/NiO/YIG shows a broad maximum at high temperatures similar to that of the enhancement of spin conductance due to AF magnons and spin fluctuation [15-17]. As found in our previous work using SSE [15], the spin conductance has a maximum near the $T_N$ of the NiO layer that increases with the NiO thickness due to the finite size effects.

The $\Delta R/R_0$ in the Pt/YIG is always positive at the measured $T$ range. Notably, there is a specific temperature $T^*$, at which the $\Delta R/R_0$ of Pt/NiO/YIG crosses zero. $T^*$ is lower than the $T_N$ of the NiO layer, and increases with the NiO thickness. At $T^*$, $R$ does not change with either the amplitude or direction of the applied field, as shown in Fig. 3(b) and 3(c) for the Pt(3)/NiO(1)/YIG at $T^* = 130$ K. As $T < T^*$, the $\Delta R/R_0$ becomes negative. This is illustrated in Fig. 3(b) for Pt(3)/NiO(1)/YIG at $T = 60$ K, with $R_\parallel < R_T$, opposite to that at $T = 300$ K (Fig. 1(c)). The angular scan (Fig. 3(c)) for the Pt(3)/NiO(1)/YIG at $T = 60$ K also shows exactly the opposite to that $T = 300$ K (Fig. 1(d)). The inverted SMR behaviors as $R_\perp \approx R_\parallel < R_T$.

From the theory of SMR for the NM/YIG structure [24], the angle-dependent MR ratio can be expressed as

$$\frac{\Delta\rho}{\rho} \approx \theta_{SH}^2 \frac{\lambda_N^2}{t_N} \frac{2G_r \tanh^2\left(\frac{t_N}{2\lambda_N}\right)}{\sigma_N + 2\lambda_N G_r \coth\left(\frac{t_N}{\lambda_N}\right)}, \quad (1)$$

where $\theta_{SH}$, $\lambda_N$, $t_N$ and $\sigma_N$ are the spin Hall angle, spin diffusion length, thickness and electrical conductivity of the NM, respectively, and $G_r$ is the real part of spin mixing conductance at the NM/YIG interface.

In Pt/YIG and Pd/YIG, $G_r$ at the NM/YIG interface is known to be barely $T$ dependent [27,28]. The $T$ dependence of spin diffusion length $\lambda_N$ gives rise to that of $\Delta R/R_0$ in NM/YIG, as noted previously [27,28]. Neglecting the small negative SMR at low temperatures for the moment, one can use Eq. (1) to calculate $G_r$ from the measured SMR, as shown in Fig. 3(d), where the $\Delta R/R_0$ is offset by $-1.2 \times 10^{-4}$ for subtracting the negative SMR, $\theta_{SH} = 0.07$, $\lambda_N$ behaviors $1/T$ from 1.5 nm at $T = 300$ K to 4 nm at $T = 10$ K, $t_N = 3$ nm, $\sigma_N = 1.2 \times 10^6/(1 + 10^{-3}T)$ $\Omega^{-1}$ m$^{-1}$. The effective $G_r$ in the Pt/NiO/YIG can be much larger than the $G_r$ in the Pt/YIG (about $1\times10^{14}$ $\Omega^{-1}$m$^{-2}$). We find that the $T$ dependence of SMR in the Pt/NiO/YIG is dominated by that of the effective $G_r$, quite different from that in Pt/YIG. The effective $G_r$ of the NiO and its interfaces to the Pt and YIG varies strongly with $T$, consistent with that we observed using SSE [15], which is due to AF magnons and spin fluctuation mediated spin current transport [15,17].

The SMR of Pt/NiO/YIG not only exhibits strong $T$ dependence but also changes sign. To address this unusual inverted SMR, we need locate its source. We use 1 nm thick Cu as an insertion layer because of its negligible spin Hall angle and MR [31]. In Fig. 4(a), we show the $T$ dependences of the SMR in Pt(3)/Cu(1)/YIG, Pt(3)/NiO(1)/Cu(1)/YIG and Pt(3)/Cu(1)/NiO(1)/YIG. Only the SMR of Pt(3)/Cu(1)/NiO(1)/YIG shows negative at low temperatures, similar to that of Pt(3)/NiO(1)/YIG. The absence of negative SMR in the Pt/NiO/Cu/YIG at low temperature reveals the crucial role of the exchange coupled NiO/YIG interface.

As $T < T_N$ of the NiO layer, spin current transmission through the NiO reduces due to less thermal magnons and spin fluctuation [15,17]. The SMR in Pt/NiO/YIG may include spin current reflection from the Pt/NiO interface in addition to that from NiO/YIG interface. Below the $T_N$ of the AF layer, exchange spring might be formed in the AF layer coupled with FM [29,30], but the NiO moments would have different angles to the YIG magnetization with angular dependence much different from the $\cos^2\alpha$ ($\cos^2\beta$) behavior that we have observed in Pt/NiO/YIG from 10 K to 300 K. There is no evidence that the rotation of the NiO moments contributes to the observed SMR. Both conventional SMR and inverted SMR indeed depend only on the magnetization orientation of YIG.

One possible mechanism of the unusual inverted SMR in the Pt/NiO/YIG at low temperature is imbedded in the SMR theory [24]. It should be noted that for both conventional SMR and inverted SMR, $R$ does not change as the field rotated in the *zx* plane, i.e. $R_\perp \approx R_\parallel$, which is the defining feature of SMR. This is due to spin current is absorbed at the interface to the FM as **M** $\perp$ **σ**. As **M** $\parallel$ **σ**, the spin current is reflected at the interface to the FM [24]. In the conventional SMR, the spin current reflection back to Pt is considered without spin-flip. After spin current reflection, the additional $\mathbf{J}_C^{ISH}$ converted by ISHE is parallel to the applied charge current $\mathbf{J}_C$, resulting in the decrease of the measured $R$, hence $R_\perp \approx R_\parallel > R_T$ [24]. This is the usual SMR, which also exists in Pt/NiO/YIG at $T > T^*$. If the spin current flowing back to the Pt from the NiO involves spin-flip, then the direction of $\mathbf{J}_C^{ISH}$ would be *opposite* to that of the $\mathbf{J}_C$, as sketched in Fig. 4(b), leading to the *increase* of the measured $R$ and thus, $R_\perp \approx R_\parallel < R_T$, the inverted SMR, as apparently occurs in Pt/NiO/YIG at $T < T^*$. We suggest that the spin-flip scattering for the spin current flowing back from the NiO to the Pt resulting in the inverted SMR at low temperatures.

In conclusion, we demonstrate that the SMR observed in the Pt/NiO/YIG heterostructures is due to magnon spin current transmitted through the thin insulating NiO layer and reflected from the NiO/YIG interface. Unlike that in Pt/YIG, the SMR in Pt/NiO/YIG shows very strong $T$ dependence dominated by that of spin conductance due to AF magnons and spin fluctuation. The SMR in Pt/NiO/YIG even reverses sign at low temperatures due to spin-flip reflection from the AF NiO exchange coupled with the YIG.

This work was supported by the SHINES, an Energy Frontier Research Center funded by the U.S. Department of Energy, Office of Science, Basic Energy Science, under Award Grant No. DE-SC0009390. W. L. was supported in part by C-SPIN, one of six centers of STARnet, a Semiconductor Research Corporation (SRC) program sponsored by MARCO and DARPA. W. L. is grateful to Qinli Ma for technical assistance and Brent Page for some calculations.

**Figure Captions**

FIG. 1. (a) X-ray diffraction of a 800 nm thick polycrystalline NiO film. (b) Schematic of angle-dependent magnetoresistance measurement in Pt/NiO/YIG. The magnetic field **H** was applied in the *xy*, *yz* and *zx* planes with angles $\alpha$, $\beta$ and $\gamma$ relative to the *x*, *z* and *x* directions, respectively. (c) $R$ of the Pt(3)/NiO(1)/YIG and Pt(3)/NiO(1)/SiO$_X$/Si at $T$ = 300 K as a function of $H$ along the *x* axis ($R_\parallel$) and the *y* axis ($R_T$), respectively. (d) Angular dependence of $R$ in Pt(3)/NiO(1)/YIG under the 0.5 T field at $T$ = 300 K. The number in the layered structure denotes thickness in nm. (e) Schematic of spin transport in the Pt/NiO/YIG. Spin current generated by the SHE in the Pt transmits through the NiO and is reflected at the NiO/YIG interface as **M** ∥ **σ**. The spin current reflected from the NiO/YIG interface can be dominated with spin current polarization along -*y*.

FIG. 2. (a) NiO thickness dependence of the SMR ratio $(R_\parallel - R_T)/R_0$ in the Pt(3)/NiO($t_{NiO}$)/YIG at room temperature. (b) CoO thickness dependence of the $(R_\parallel - R_T)/R_0$ in the W(3)/CoO($t_{CoO}$)/YIG at room temperature. $R_\parallel$ and $R_T$ were measured at the 0.5 T field, and $R_0$ was measured at zero field.

FIG. 3. (a) Temperature dependences of the SMR ratio in the Pt(3)/YIG, Pt(3)/NiO(0.6)/YIG, Pt(3)/NiO(1)/YIG and Pt(3)/NiO(2)/YIG at the 0.5 T field. (b) $R$ of the Pt(3)/NiO(1)/YIG at $T$ = 130 K and 60 K as a function of $H$ along the *x* axis ($R_\parallel$) and the *y* axis ($R_T$), respectively. (c) Angular dependence of $R$ in Pt(3)/NiO(1)/YIG under the 0.5 T field at $T$ = 130 K and 60 K. (d) Deduced effective $G_r$ as a function of $T$ in the Pt(3)/NiO(0.6)/YIG, Pt(3)/NiO(1)/YIG and Pt(3)/NiO(2)/YIG from the measured SMR ratio with subtracting the negative SMR.

FIG. 4. (a) Temperature dependences of the SMR ratio in the Pt(3)/Cu(1)/YIG, Pt(3)/NiO(1)/Cu(1)/YIG and Pt(3)/Cu(1)/NiO(1)/YIG at the 0.5 T field. (b) Schematic of spin transport in the Pt/NiO/YIG as $T < T^*$. Spin current generated by the SHE in the Pt transmits through the NiO and is reflected from the NiO as **M** ∥ **σ**. The spin current flowing back from the NiO to the Pt can be dominated with spin current polarization along +*y* as $T < T^*$.

**Figures**

FIG. 1.

FIG. 2

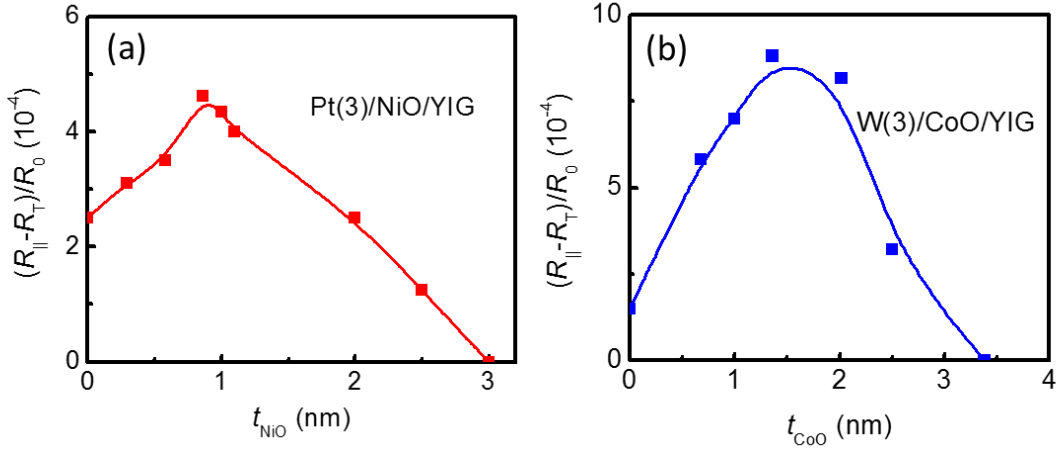

FIG. 3

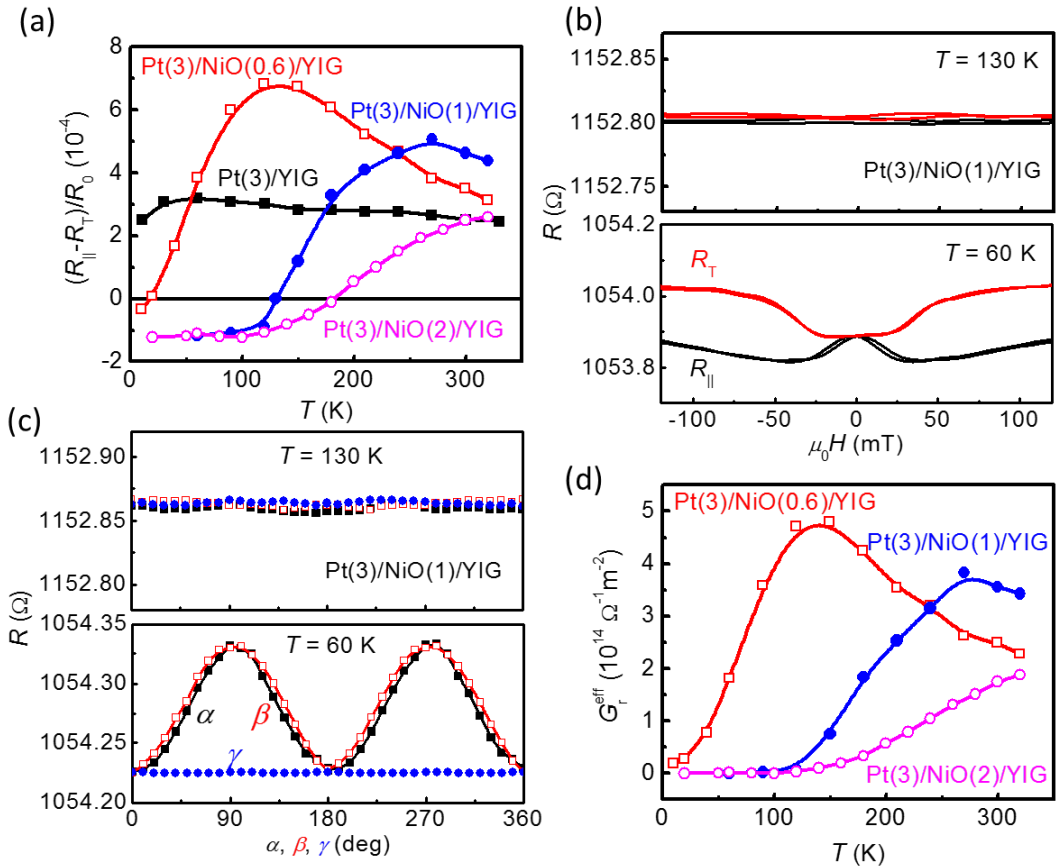

FIG. 4

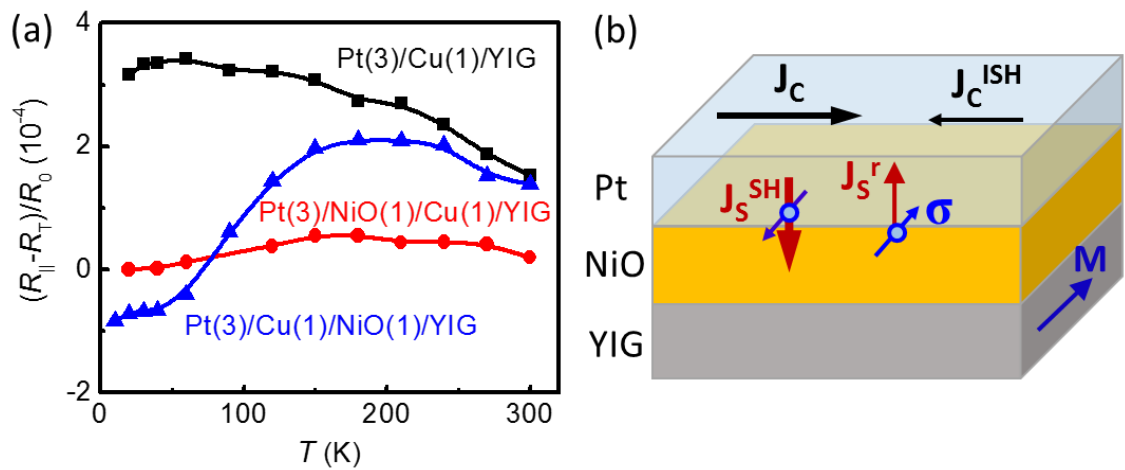